\documentclass[twocolumn,showpacs,preprintnumbers,amsmath,amssymb]{revtex4}
\usepackage{graphicx}% Include figure files                                     
\usepackage{dcolumn}% Align table columns on decimal point                      
\usepackage{bm}% bold math                                                      
\begin{document}
%\preprint{APS/123-QED}
\title{Pseudo-Quantum Criticality in Electron Liquids Exhibited 
in Expanded Alkali Metals}% Force line breaks with \\
\author{Hideaki Maebashi and Yasutami Takada}
\affiliation{Institute for Solid State Physics, 
University of Tokyo, Kashiwa, Chiba 277-8581, Japan}
\date{\today}% It is always \today, today,
\begin{abstract}
With paying special attention to the divergence in the compressibility $\kappa$, 
we study the Coulombic screening in alkali metals to find singular long-range 
fluctuations in the electronic polarization originating from this divergence. 
As a consequence of this singularity, we predict the decrease of the 
equilibration distance between ions against the increase of $r_s$ the 
Wigner-Seitz radius of valence electrons, provided that the condition of 
$2r_{\rm c} < r_s <4r_{\rm c}$ is satisfied with $r_{\rm c}$ the ion-core radius. 
This prediction is in good quantitative agreement with the recent experiment 
on liquid Rb. 
\end{abstract}

\pacs{71.10.Ca, 71.45.Gm, 61.25.Mv, 05.70.Jk}
% PACS, the Physics and Astronomy Classification Scheme.                        
\maketitle

%%%%%%%%% Paragraph 1: Introduction; relate with experiment on Rb  %%%%%%%%%%%%%
According to the recent high-resolution measurement of x-ray diffraction 
for expanded liquid Rb \cite{MTI07}, an average distance between 
nearest-neighbor ions $R_{\rm NN}$, as determined by the first-peak position 
in the radial distribution function $g(R)$, {\it decreases} with increasing 
$R_0$ the Wigner-Seitz (WS) radius of ions. This is in sharp contrast with 
the normal situation in compressed liquid alkali metals in which $R_{\rm NN}$ 
increases with $R_0$ \cite{Tsuji96,FLHM05,GDSHM05,Shimojo97}. In solid Rb the 
valence-electron density $n$ is specified by $r_s=5.2$ with $r_s$ being 
the WS radius of electrons (in atomic units which we take throughout this 
Letter), very close to $r_s=5.25$ at which the compressibility $\kappa$ 
of the 3D electron gas diverges \cite{CA80}. Thus $n$ in expanded liquid Rb is 
in the range of $r_s>5.25$ where $\kappa$ is negative \cite{comment1}. In this Letter, 
we shall explain this unexpected contraction of $R_{\rm NN}$ by focusing on 
the divergence, rather than the negativeness, of $\kappa$ as a key issue. 

%%%%%%%%% Paragraph 2: Introduction; discuss peaudocriticality     %%%%%%%%%%%%%
At arbitrary $n$, the compressibility sum rule relates $\kappa$ to the 
polarization function $\Pi^{R}(q, \omega)$ (with the superscript $R$ denoting 
the retarded form) through $\lim_{q \to 0}\Pi^{R}(q, 0)=n^2 \kappa$. Hence, 
near the critical density of $r_s=5.25$, $\Pi^{R}(q, \omega)$ becomes singular 
in the limits of $q/p_{\rm F} \to 0$ and $\omega/v_{\rm F}q \to 0$ with 
$p_{\rm F}$ and $v_{\rm F}$ being, respectively, the Fermi wave number and the 
free-electron Fermi velocity. More specifically, we find that 
$\Pi^{R}(q,\omega)$ is well characterized by such a singular form as 
\begin{equation}
\Pi^{R}(q, \omega) \simeq n^2 \kappa_{\rm F} \xi_{\rm F}^{-2} 
/(\xi^{-2} + q^2 - {\rm i} \omega / \Gamma q)\,,
\label{PQCP}
\end{equation}
where $\xi^2=(\kappa/\kappa_{\rm F})\xi_{\rm F}^2$, $\Gamma=(2/\pi)v_{\rm F}
\xi_{\rm F}^2$, and $\xi_{\rm F}^{-2}=12p_{\rm F}^2$ with $\kappa_{\rm F}$ 
the free-electron compressibility. Note that near the quantum critical 
point (QCP) of dynamical exponent $3$ \cite{Sachdev99}, the response function 
takes exactly the same form as in Eq.~(\ref{PQCP}). In the electron gas, 
however, there is no true phase transition (and in fact ``the correlation 
length'' $\xi$ turns into a pure imaginary value for negative $\kappa$), owing 
to the fact that $\Pi^{R}(q, \omega)$ in this case represents response to a 
total of external and induced charges, making it acausal \cite{DKM81}. One of 
the interesting consequences of Eq.~(\ref{PQCP}) near this ``pseudo-quantum 
critical point'' is the softening of excitonic collective modes, as featured 
by the dispersion relation of $\omega=(2/\pi)|\, \kappa_{\rm F}/\kappa+q^2 
\xi_{\rm F}^2\,|v_{\rm F}q$ \cite{Takada01}. 

%%%%%%%%% Paragraph 3: Introduction; scope of the paper            %%%%%%%%%%%%%
The classical or Coulombic screening of ions near this pseudo-QCP poses a 
distinct problem from the quantum or Kondo screening near the true 
QCP \cite{MMV02} and deserves special attention, basically because it is useful 
in clarifying the important physics of the coupling between the $q=0$ 
singularity in the Coulomb interaction, $v(q)=4\pi/q^2$, and the singular 
fluctuations in $\Pi^{R}(q, 0)$. Another important aspect of the problem is the 
fact that the ions in alkali metals are not point charges but complexes of a 
nucleus and core electrons surrounding it, the feature of which can be well 
captured by a suitable pseudopotential $V_{\rm ps}(q)$. This complex structure 
combined with the pseudo-QCP makes the observable phenomena intriguing, in 
which an important role is played by the presence of a characteristic wave 
number $q_0$ (proportional to the inverse of the core-electron radius 
$r_{\rm c}$) at which $V_{\rm ps}(q)$ vanishes due to the orthogonality between 
valence- and core-electron wave functions \cite{Harrison89}. 

\begin{figure}[t]
\includegraphics[width=8.4cm]{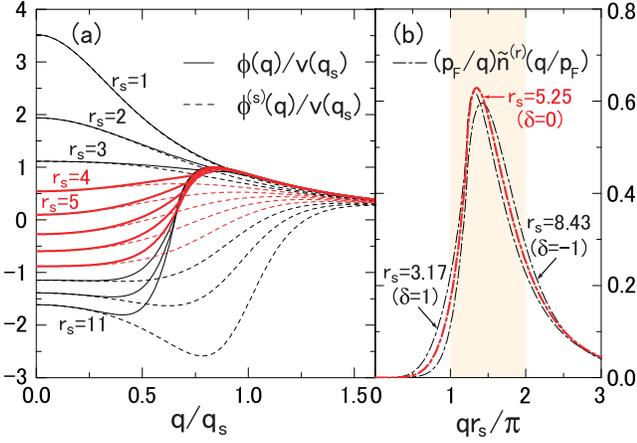}% Here is how to import EPS art
\caption{
(a) Electrostatic potential $\phi(q)$ scaled by $v(q_{\rm s})$ with $q_{\rm s}
=\sqrt{q_{\rm TF}/\xi_{\rm F}}$ as a function of $q/q_{\rm s}$ with changing 
$r_s$ from $1$ to $11$ by $1$ (the solid curves from top to bottom), 
together with the corresponding screening potential $\phi^{({\rm s})}(q)$ 
in Eq.~(\ref{electrostatic potential}) by the broken curves. 
(b) Rearrangement of the charge distribution due to the screening hole 
${\tilde n}^{({\rm r})}(q/p_{\rm F})$ divided by $q/p_{\rm F}$, which is 
properly scaled in Eq.~(\ref{universal}), as a function of $q r_s / \pi$.  
The ``distance'' from the pseudo-QCP is defined as 
$\delta=\phi(0)/v(q_{\rm s})$.}
\label{fig:1}
\end{figure}

%%%%%%%%% Paragraph 4: Introduction; summary of the results        %%%%%%%%%%%%%
In this Letter, we report on our finding that the pseudo-quantum criticality 
is intimately connected with the appearence of a local minimum in the 
electrostatic potential $\phi(q)$ at $q=0$ for some range of $r_s$ (see 
Fig.~\ref{fig:1}). This minimum is shown to be brought about by (i) the 
Thomas-Fermi-type screening with the singular polarization and (ii) the 
subsequent rearrangement of the charge distribution in response to the resulting 
screening hole. The first process induces the contraction of the equilibration 
distance between ions, while the second acts in the opposite way. In expanded 
alkali metals, this second process is suppressed by the presence of $q_0$ in 
$V_{\rm ps}(q)$, leading to the experimental results~\cite{MTI07} with which 
our theoretical results of $g(R)$ obtained by systematic Monte-Carlo 
simulations for Rb ions are in good accord. 

%%%%%%%%%%%%%%%% Paragraph 5: Electrostatic potential        %%%%%%%%%%%%%%%%%%%%
In terms of $\Pi^{R}(q,0)$, we can write $\phi(q)$ induced by a point charge 
in the 3D electron gas as 
\begin{equation}
\phi (q) = v(q) / [\, 1 + v (q) \Pi^{R} (q,0) \,]\, .
\label{effective interaction}
\end{equation}
Using the exchange-correlation kernel $f_{\rm xc}(q, \omega)$, we can give 
a formally exact experession for $\Pi^{R}(q,\omega)$ as $\Pi^{R}(q,\omega)=
[\Pi^{0R}(q,\omega)^{-1}+f_{\rm xc}(q, \omega)]^{-1}$, where 
$\Pi^{0R}(q,\omega)$ is the free-electron polarization function. 
Since an accurate parameterization of the diffusion Monte Carlo (DMC) data 
for $f_{\rm xc }(q,0)$ is available \cite{MCS95}, the virtually exact 
$\phi (q)$ is known. (See the solid curves in Fig.~\ref{fig:1}(a).) 
The DMC data reveal that $f_{\rm xc} (q, 0)$ is flat near $q=0$, e.g., 
$f_{\rm xc}(q,0)-\lim_{q \to 0}f_{\rm xc}(q, 0)$ $\sim$ $O(q^{16})$ for 
$r_s = 5$ \cite{MCS95}, allowing us to derive the result in Eq.~(\ref{PQCP}) 
for $q/p_{\rm F} \ll 1$ and $\omega/v_{\rm F}q \ll 1$ under the assumption 
that the limits of $q \to 0$ and $\omega \to 0$ are interchangeable for 
$f_{\rm xc}(q,\omega)$. 

%%%%%%%%%%%%%%%% Paragraph 6: Length Scale       %%%%%%%%%%%%%%%%%%%%
Although $\Pi^{0R} (q,0)$ itself is specified only by $p_{\rm F}^{-1}=
\alpha r_s$ with $\alpha=(4/9 \pi)^{1/3}$, a unique length scale for this 
function, consideration of the coupling to the Coulomb interaction evokes 
another length scale, $q_{\rm TF}^{-1}=(\pi \alpha r_s/4)^{1/2}$, as 
characterized by $v(q)\Pi^{0R}(q,0) \simeq q_{\rm TF}^2/q^2$ for $q \to 0$. 
For the full polarization function $\Pi^{R} (q,0)$, Eq.~(\ref{PQCP}) 
indicates $|\xi|$ as a characteristic length scale which becomes much 
longer than $\xi_{\rm F} \approx p_{\rm F}^{-1}$ for $\kappa/\kappa_{\rm F} 
\gg 1$. Eventually at the pseudo-QCP, concomitantly with the divergence 
of $|\xi|$, a new length scale, $q_{\rm s}^{-1}=\sqrt{\xi_{\rm F}/q_{\rm TF}}$, 
emerges for the coupling to the Coulomb interaction, as characterized by 
$v(q) \Pi^{R}(q,0) \simeq q_{\rm s}^4/q^4$ for $q \to 0$. 

%%%%%%%%%%%%%%%% Paragraph 7: Local minimum in \phi(q)   %%%%%%%%%%%%%%%%%%%%
Keeping those length scales in mind, we represent $\phi(q)$ in an expansion 
form around $q=0$ for arbitrary $n$ as 
\begin{equation}
\phi(q) = v(q_{\rm s}) [ \, \delta + (1 - \delta^2) 
(q/q_{\rm s})^2 + O(q^4) \, ] \, .
\label{expansion}
\end{equation}
Here we have introduced $\delta$ $[\equiv (\kappa_{\rm F}/\kappa)/(q_{\rm TF}
\xi_{\rm F})]$ as a key quantity describing the ``distance'' from the 
pseudo-QCP. In Eq.~(\ref{expansion}), the coefficient of the $q^2$ term is 
positive for $|\delta|<1$ corresponding to $3.17<r_s<8.43$, which clarifies 
the reason why $\phi(q)$ in Fig.~\ref{fig:1}(a) exhibits a local minimum at 
$q=0$ only in this range of $r_s$. We shall call this range the pseudo-quantum 
critical region, in which alkali metals, Na, K, Rb, and Cs, are included. 

%%%%%%%%%%%%%%%%%%%%%%%%%%%%%%% Figure 1:  %%%%%%%%%%%%%%%%%%%%%%%%%%%%%%%%%%%%%

%%%%%%%%% Paragraph 8: Extracting the singular screening potential %%%%%%%%%%%%%
In order to obtain a deeper insight into the Coulombic screening in the 
presence of the singular contribution in $\Pi^{R}(q,0)$, let us write 
$\phi(q)$ as
\begin{equation}
\phi (q) = v(q) [\, 1 - n_{\rm ind}(q) \, ] = \phi^{({\rm s})}(q) 
+ \phi^{({\rm r})}(q) \,,
\label{electrostatic potential}
\end{equation}
where $n_{\rm ind}(q)$ $[\equiv v(q)\Pi^{R}(q,0)/(1+v(q)\Pi^{R}(q,0))]$ 
is the electron density distribution induced by the point charge and 
$\phi^{({\rm s})}(q)$ is defined by substituting the singular polarization, 
Eq.~(\ref{PQCP}), into Eq.~(\ref{effective interaction}) as 
$\phi^{({\rm s})}(q)=v(q)[1-n_{\rm ind}^{({\rm s})}(q)]$ with 
$n_{\rm ind}^{({\rm s})}(q)$, given by
\begin{equation}
n_{\rm ind}^{({\rm s})}(q) = [ \, 1+\delta\, (q/q_{\rm s})^2 + 
(q/q_{\rm s})^4 \, ]^{-1} \,.
\label{screening charge}
\end{equation}
It is easily verified that when $r_s \to 0$, Eq.~(\ref{screening charge}) 
is reduced to the standard Thomas-Fermi screening charge, 
$n_{\rm ind}^{({\rm s})}(q) \to q_{\rm TF}^2/(q_{\rm TF}^2+q^2)$. Thus 
we may regard $\phi^{({\rm s})}(q)$ as the Thomas-Fermi-type screening 
potential with the singular polarization. In Fig.~\ref{fig:1}(a), 
$\phi^{({\rm s})}(q)$ is plotted by the broken curves with changing 
$r_s$ from $1$ to $11$ by $1$. 

%%%%%%%%%%%%%%%% Paragraph 9: Residual potential  %%%%%%%%%%%%%%%%%%%%
The residual term in Eq.~(\ref{electrostatic potential}), 
$\phi^{({\rm r})}(q)$, is the electrostatic potential due to the rearrangement 
of the induced charge from $n_{\rm ind}^{({\rm s})}(q)$ to $n_{\rm ind}(q)$, 
amounting to $\phi^{({\rm r})}(q)=v(q)n^{({\rm r})}(q)$ with $n^{({\rm r})}(q) 
\equiv n_{\rm ind}^{({\rm s})}(q)-n_{\rm ind}(q)$. Since the singular 
long-range part $n_{\rm ind}^{({\rm s})}(q)$ is extracted from 
$n_{\rm ind}(q)$, we speculate that $n^{({\rm r})}(q)$ is characterized 
only by the length scale $p_{\rm F}^{-1}$, implying that $n^{({\rm r})}(q)$ 
should be expressed in terms of some universal function of $q/p_{\rm F}$. 
This speculation is checked by casting $n^{({\rm r})}(q)$ into the form as 
\begin{equation}
n^{({\rm r})}(q) = (2 + \delta)^{-1}(q_{\rm TF}/q_{\rm s})^2\,
{\tilde n}^{({\rm r})}(q/p_{\rm F})\,.
\label{universal}
\end{equation}
Here we have determined the prefactor by respecting the result of 
$n_{\rm ind}^{({\rm s})}(q_{\rm s})=(2+\delta)^{-1}$. In Fig.~\ref{fig:1}(b), 
we plot the obtained ${\tilde n}^{({\rm r})}(x)/x$ with $x=q/p_{\rm F}$ for 
$\delta =0$ and $\pm 1$, from which we see that ${\tilde n}^{({\rm r})}(x)/x$, 
having a peak structure for $\pi/r_s<q<2\pi/r_s$, may well be regarded as 
universal in the entire pseudo-quantum critical region. 

%%%%%%%%%%%%%%% Paragraph 10:  Formula for Ion-ion interaction  %%%%%%%%%%%%%%%%
In calculating ${\tilde V}_{\rm ii} (R)$ the effective interaction 
between ions, we resort to second-order perturbation with respect to 
$V_{\rm ps}(q)$ \cite{AS78}. For $R$ outside the core region, we obtain 
\begin{equation}
{\tilde V}_{\rm ii}(R) = \frac{1}{2 \pi^2 R} 
\int_{0}^{\infty} \rho_{\rm ps}(q)^2 \phi(q) q \sin(qR) {\rm d} q \,,
\label{ion-ion interaction}
\end{equation}
where $\rho_{\rm ps}(q)$ is the effective charge of the ion, given by 
$\rho_{\rm ps}(q)=(q^2/4 \pi)V_{\rm ps}(q)$. In general, the charge 
neutrality of the whole system imposes $\rho_{\rm ps}(0)=1$. For $r_{\rm c}>0$, 
$\rho_{\rm ps}(q)$ vanishes at $q=q_0$. In the limit of $r_{\rm c} \to 0$ 
(or $q_0 \to \infty$), however, $\rho_{\rm ps}(q)=1$ for arbitrary $q$, which 
reduces ${\tilde V}_{\rm ii} (R)$ into $\phi (R)$, the inverse Fourier 
transform of $\phi (q)$. In the following specific calculations, for 
simplicity, we employ the Ashcroft empty-core pseudopotential, for which 
$\rho_{\rm ps}(q)=\cos q r_{\rm c}$, but our conclusions do not depend 
on the detailed shape of $V_{\rm ps}(q)$ very much. 

%%%%%%%%%%%%%%%% Figure 2         %%%%%%%%%%%%%%%%%%%%
\begin{figure}[lt]
\includegraphics[width=8.3cm]{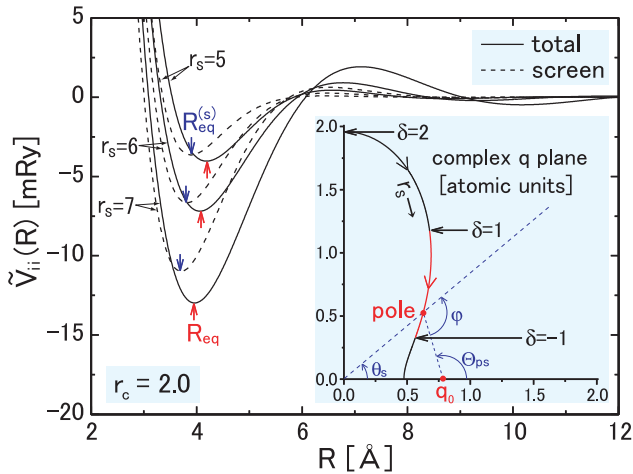}% Here is how to import EPS art
\caption{
Effective interaction between ions, ${\tilde V}_{\rm ii}(R)$, 
with using the empty-core pseudopotential with $r_{\rm c} = 2.0$, for 
$r_s = 5$, $6$, and $7$ (the solid curves). 
The broken curves represent ${\tilde V}_{\rm ii}^{({\rm s})}(R)$ 
in Eq.~(\ref{screened interaction}). The inset shows the trajectory 
of the screening pole $q_{\rm s}{\rm e}^{{\rm i}\theta_{\rm s}}$ 
with increasing $r_s$ in the complex $q$ plane.}
\label{fig:2}
\end{figure}

%%%%%%%%%%%%%%%% Paragraph 11: Analytical structure        %%%%%%%%%%%%%%%%%%%%
In accordance with Eq.~(\ref{electrostatic potential}), we shall devide 
${\tilde V}_{\rm ii} (R)$ in Eq.~(\ref{ion-ion interaction}) into two 
components as 
\begin{equation}
{\tilde V}_{\rm ii}(R) = {\tilde V}_{\rm ii}^{({\rm s})} (R) 
+ {\tilde V}_{\rm ii}^{({\rm r})} (R) \,,
\label{two contributions in V}
\end{equation}
where ${\tilde V}_{\rm ii}^{({\rm s})}(R)$ and 
${\tilde V}_{\rm ii}^{({\rm r})}(R)$ are, respectively, defined 
by using $\phi^{({\rm s})}(q)$ and $\phi^{({\rm r})}(q)$ instead of 
$\phi(q)$ in Eq.~(\ref{ion-ion interaction}). Since 
$n_{\rm ind}^{({\rm s})}(q)$ is given in a simple analytic function, we can 
analyze ${\tilde V}_{\rm ii}^{({\rm s})}(R)$ by pursuing the poles of 
$\phi^{({\rm s})}(q)$ or equivalently those of $n_{\rm ind}^{({\rm s})}(q)$ 
in the complex $q$ space. According to Eq.~(\ref{screening charge}), all the 
poles are on the imaginary axis for $\delta>2$, while they deviate from the 
axis for $\delta<2$ (see the inset of Fig.~\ref{fig:2}), implying that 
the ``screening length" in the Thomas-Fermi-type screening potential 
changes into a complex number, so that the resulting potential shows 
a damped oscillatory behavior. As $\delta$ decreases, the oscillation 
amplitude increases and eventually for $\delta<1$, the short-range 
attractive part develops in ${\tilde V}_{\rm ii}^{({\rm s})}(R)$. 
Note that this oscillation related to the pseudo-quantum criticality is 
distinct from the Friedel oscillation, because the latter does not originate 
from those poles but the branch-cut singularity in $\phi(q)$. 

%%%%%%%%%%%%%%%% Paragraph 12: Discuss the pole contribution    %%%%%%%%%%%%%%%%
For $|\delta|<2$, the poles in the upper-half complex $q$ plane reside at 
$q_{\rm s}{\rm e}^{{\rm i}\theta_{\rm s}}$ and $-q_{\rm s}{\rm e}^{-{\rm i}
\theta_{\rm s}}$ with $\theta_{\rm s}=(1/2)\cos^{-1}(-\delta/2)$. Writing 
$\rho_{\rm ps}(q_{\rm s}{\rm e}^{{\rm i}\theta_{\rm s}})$ as 
$Z_{\rm ps}{\rm e}^{{\rm i} \Theta_{\rm ps}}$, we obtain 
${\tilde V}_{\rm ii}^{({\rm s})}(R)$ as 
\begin{equation}
{\tilde V}_{\rm ii}^{({\rm s})}(R) = 
Z_{\rm ps}^2 \frac{\sin (2\varphi -q_{\rm s}R\cos \theta_{\rm s})}
{R\sin 2 \theta_{\rm s}}\ {\rm e}^{- q_{\rm s} R \sin \theta_{\rm s}}\,,
\label{screened interaction}
\end{equation}
with $\varphi=\theta_{\rm s}-\Theta_{\rm ps}+\pi$. This analytic result 
with $r_{\rm c}=2.0$ is plotted in Fig.~\ref{fig:2} by the broken curves for 
$r_s=5$, 6, and 7, in comparison with the corresponding numerical result for 
${\tilde V}_{\rm ii}(R)$ (the solid curves). The global minimum in 
${\tilde V}_{\rm ii}(R)$ determines the equilibration distance between ions, 
$R_{\rm eq}$, which is found to decrease with increasing $r_s$. The same is 
true for $R_{\rm eq}^{({\rm s})}$ the equilibration distance determined 
through ${\tilde V}_{\rm ii}^{({\rm s})}(R)$, suggesting that the main feature 
of $R_{\rm eq}$ will be studied by examining $R_{\rm eq}^{({\rm s})}$. 
An approximate result for $R_{\rm eq}^{({\rm s})}$ is obtained from 
Eq.~(\ref{screened interaction}) as
\begin{equation}
R_{\rm eq}^{({\rm s})} \approx (2\varphi+\pi/2)/(q_s\cos \theta_{\rm s})
= (4\varphi + \pi)/q_s\sqrt{2-\delta}\,.
\label{equilibrium distance}
\end{equation}
A rather extensive examination of various factors in 
Eq.~(\ref{screened interaction}) reveals that $R_{\rm eq}^{({\rm s})}$ is 
mainly controlled by $\varphi$. As shown in the inset of Fig.~\ref{fig:2}, 
the phase $\theta_{\rm s}$ decreases as the pole $q_{\rm s}{\rm e}^{{\rm i}
\theta_{\rm s}}$ evolves along the trajectory with increasing $r_s$. On the 
other hand, the phase $\Theta_{\rm ps}$ increases, because it is approximately 
given by the phase of $q-q_0$ itself, as can be seen by expanding 
$\rho_{\rm ps}(q)$ around $q=q_0$. Therefore, in a combined manner, both 
changes contribute to the decrease of the phase $\varphi$, leading eventually 
to the contraction of $R_{\rm eq}^{({\rm s})}$ with the increase of $r_s$. 
Incidentally, for very large $q_0$ as in the case of a point charge, 
$\Theta_{\rm ps}$ hardly changes. Thus in such a case, only the phase 
$\theta_{\rm s}$ contributes to the decrease of $R_{\rm eq}^{({\rm s})}$. 

%%%%%%%%%%%%%%%% Paragraph 13: Contribution from the residual charge   %%%%%%%%%
The presence of ${\tilde V}_{\rm ii}^{({\rm r})}(R)$ in 
Eq.~(\ref{two contributions in V}) manifests itself in the difference between 
$R_{\rm eq}$ and $R_{\rm eq}^{({\rm s})}$. This residual interaction is 
determined by $n^{({\rm r})}(q)$ in Eq.~(\ref{universal}). Since it 
is scaled by $p_{\rm F}^{-1}$, the resulting interaction is necessarily 
scaled by the same length scale or $\alpha r_s$, indicating that 
${\tilde V}_{\rm ii}^{({\rm r})}(R)$ 
contributes to increasing $R_{\rm eq}^{({\rm s})}$ with $r_s$. 
In this context, it is important to note that $n^{({\rm r})}(q)$ 
has a sharp peak structure as shown in Fig.\ref{fig:1} (b). Thus, 
if $\rho_{\rm ps}(q)$ is small in the peak region of $n^{({\rm r})}(q)$ 
as realized by the condition of $\pi/r_s<q_0<2\pi/r_s$, the contribution of 
${\tilde V}_{\rm ii}^{({\rm r})}(R)$ is strongly suppressed, making 
$R_{\rm eq}$ approach $R_{\rm eq}^{({\rm s})}$. This suppression never 
occurs in the point-charge case.

%%%%%%%%%%%%%%%% Figure 3         %%%%%%%%%%%%%%%%%%%%
\begin{figure}[rb]
\includegraphics[width=7.6cm]{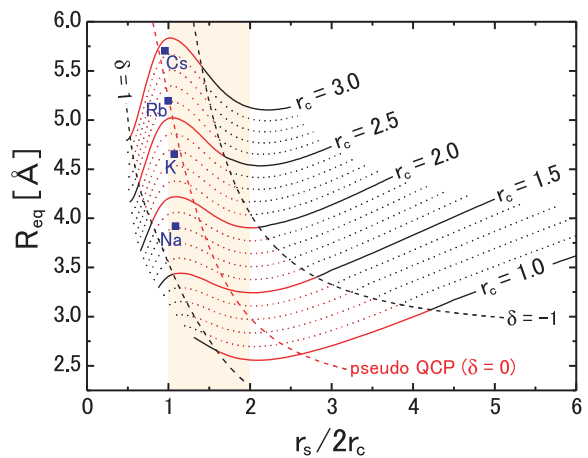}% Here is how to import EPS art
\caption{Equilibration distance between ions, $R_{\rm eq}$, for the 
empty-core pseudopotential with changing $r_{\rm c}$ from 
$1.0$ to $3.0$ by $0.1$ versus the ratio of the WS radius of the 
electrons $r_s$ to the core diameter $2 r_{\rm c}$. }
\label{fig:3}
\end{figure}

%%%%%%%%%%%%%%%% Figure 4         %%%%%%%%%%%%%%%%%%%%
\begin{figure}[t]
\includegraphics[width=7.8cm]{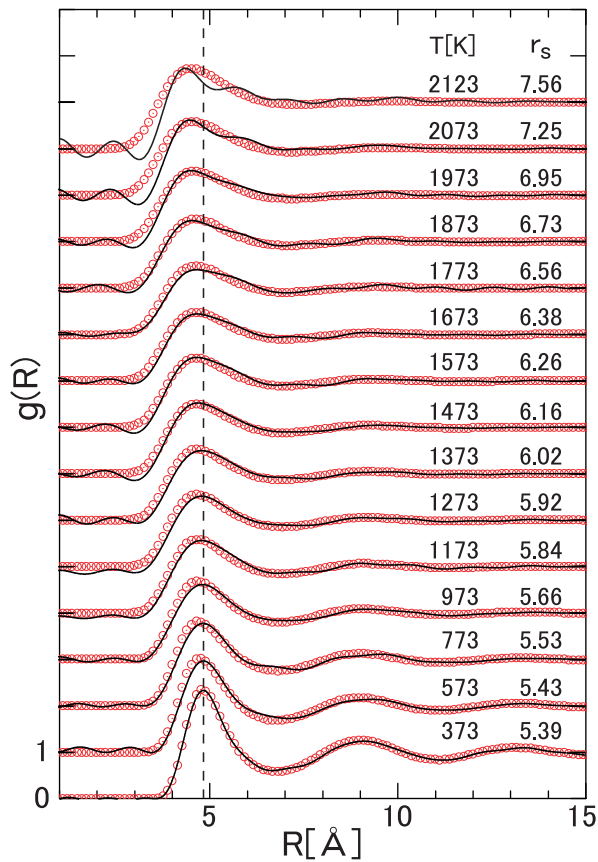}% Here is how to import EPS art
\caption{Comparison between our theoretical results of $g(R)$ obtained 
by the canonical Monte-Carlo simulations (the open circles) and 
the experimental results in Ref.~[1] (the solid curves). 
The broken line denotes $R_{\rm NN}$ of $g(R)$ at $373$ K.}
\label{fig:4}
\end{figure}

%%%%%%%%%%%%%%%% Paragraph 14:   Overall structure         %%%%%%%%%%%%%%%%%%%%
In Fig.~\ref{fig:3}, we show the overall behavior of $R_{\rm eq}$ with the 
change of $r_s/2r_{\rm c}$, which is obtained numerically with using 
the empty-core pseudopotential. 
The broken lines correspond to $\delta = 1$ ($r_s = 3.17$), 
$\delta = 0$ ($r_s = 5.25$), and $\delta = -1$ ($r_s = 8.43$), 
showing the pseudo-quantum critical region. The contraction of $R_{\rm eq}$ 
is clearly seen for $2r_{\rm c}<r_s<4r_{\rm c}$ in accordance with 
$\pi/r_s<q_0<2\pi/r_s$, where $q_0=\pi/2r_{\rm c}$. Although 
the global minimum of ${\tilde V}_{\rm ii}(R)$ is obtained at its first 
minimum for $\delta < 1$, the second minimum becomes lower than the first 
one for $\delta>1$, making $R_{\rm eq}$ jump, which is not shown in 
Fig.~\ref{fig:3}. For $\delta<-1$, the situation is completely different 
because of the shift of the global minimum of $\phi (q)$ to a finite $q$ 
(see Fig.~\ref{fig:1}). This may be related to the instability 
toward the Wigner transition. 

%%%%%%%%%%%%%%%% Paragraph 15:         %%%%%%%%%%%%%%%%%%%%
The contraction of $R_{\rm eq}$ may be explained from a geometrical point of 
view. Inserting two ions into the electron liquid excludes the valence 
electrons from the core regions owing to the orthogonality between 
valence- and core-electron wave functions. For the case of 
$2r_{\rm c}<r_s<4r_{\rm c}$, the rearrangement of the induced charge 
distribution due to the screening hole is restricted, because the WS radius 
$r_s$, which is the typical length of the electron liquid, is comparable with 
the diameter of the excluded volume. Then, ${\tilde V}_{\rm ii}(R)$ is dominated 
by the complex screening ${\tilde V}_{\rm ii}^{({\rm s})}(R)$ in the 
pseudo-quantum critical region. On the other hand, when $r_s>4r_{\rm c}$, 
the valence electrons recognize the ion core virtually as a point charge, 
so that the problem is reduced to $R_{\rm eq}$ between two point charges 
which is an increasing function of $r_s$. In Fig.~\ref{fig:3}, we have also 
plotted $R_{\rm eq}$ for solid alkali metals, Na, K, Rb, and Cs. 
Since $r_s/2r_{\rm c} \simeq 1$ in these metals, the contraction of 
$R_{\rm eq}$ is generally expected in the expanded alkali metals.

%%%%%%%%%%%%%%%% Paragraph 16:  Comparison with experiment  %%%%%%%%%%%%%%%%%%%%
Finally, in Fig.~\ref{fig:4}, we present the canonical Monte-Carlo results of 
the radial distribution function $g(R)$ for $500$ classical particles 
interacting with each other through ${\tilde V}_{\rm ii} (R)$ in 
Eq.~(\ref{ion-ion interaction}) at finite temperatures, together with the 
experimental results in Ref.~\cite{MTI07}. For liquid Rb, we have used the 
empty-core pseudopotential with $r_{\rm c}=2.4$ \cite{TCH95}. The agreement 
between theory and experiment is good not only for the first-peak position 
but also for the overall change of $g(R)$ with increasing $r_s$ and temperature. 

%%%%%%%%%%%%%%%% Paragraph 17:   Summary      %%%%%%%%%%%%%%%%%%%%
In summary, we have shown that the pseudo-quantum criticality in electron 
liquids transforms the standard Thomas-Fermi screening into the complex 
screening characterized by the development of a short-range attractive part 
in the screened Coulomb interaction. For $2r_{\rm c}<r_s<4r_{\rm c}$, 
the core-valence orthogonality promotes this complex-screening effect by 
suppressing the rearrangement of the induced charge due to the screening hole, 
leading to the contraction of the equilibration distance between ions with 
increasing $r_s$ in good agreement with experimemt on expanded liquid Rb. 

%%%%%%%%%%%%%%%% Paragraph 18: Acknowledgements       %%%%%%%%%%%%%%%%%%%%
The authors acknowledge K. Matsuda and K. Tamura for providing their 
experimental data. The authors also thank T. Kato, M. Shimomoto, and 
M. Shiroishi for helpful comments on the Monte-Carlo simulations. The 
computational work is done using the facilities of Supercomputer Center, 
ISSP, UT. This work is partially supported by a Grant-in-Aid for Scientific 
Research in Priority Areas (No.17064004) of MEXT, Japan.

%%%%%%%%%%%%%%%%%%%%%%%%%%%%%%%%% References  %%%%%%%%%%%%%%%%%%%%%%%%%%%%%%%%%%
%\begin{references}


\begin{thebibliography}{99}
\bibitem{comment1}It is noted that the valence-electron system is stable 
against phase separation even though $\kappa<0$, because the total 
compressibility of electrons and ions is positive.
\end{thebibliography}
\end{document}